\pdfoutput=1

\documentclass[11pt]{article}

\usepackage[preprint]{acl}

\usepackage{longtable}
\usepackage{anyfontsize}
\usepackage{multicol}
\usepackage{lmodern}
\usepackage{relsize}
\usepackage{multirow}
\usepackage{afterpage}
\usepackage{times}
\usepackage{latexsym}
\usepackage{amssymb}
\usepackage{pifont}
\usepackage{tikz}
\usepackage{comment}
\usepackage{tcolorbox}
\tcbuselibrary{listingsutf8}

\usetikzlibrary{shapes, arrows.meta, positioning, backgrounds}

\newcommand{\cmark}{\ding{51}}%
\newcommand{\xmark}{\ding{55}}%

\usepackage[T1]{fontenc}

\usepackage[utf8]{inputenc}

\usepackage{microtype}
\usepackage{amsmath} 
\definecolor{darkgreen}{rgb}{0.0, 0.7, 0.0}

\usepackage{inconsolata}

\usepackage{graphicx}
\usepackage{xcolor}
\usepackage{booktabs}

\title{\raisebox{-0.1cm}{\includegraphics[width=0.8cm,height=0.8cm,keepaspectratio]{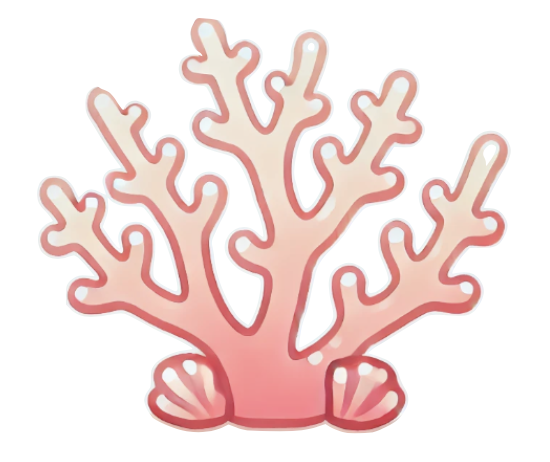}}CORAL: Benchmarking Multi-turn Conversational Retrieval-Augmentation Generation}

\author{Yiruo Cheng$^1$, Kelong Mao$^1$, Ziliang Zhao$^1$, Guanting Dong$^1$, Hongjin Qian$^2$, \\
{\bf Yongkang Wu$^3$,} {\bf Tetsuya Sakai$^4$,} {\bf Ji-Rong Wen$^1$,} {\bf Zhicheng Dou$^1$}\thanks{Corresponding author.}
\\
$^1$Gaoling School of Artificial Intelligence, Renmin University of China \\
$^2$Beijing Academy of Artificial Intelligence \\
$^3$Huawei Poisson Lab\\
$^4$Waseda University, Tokyo, Japan \\
\texttt{\{chengyr,mkl,dou\}@ruc.edu.cn}  }

\begin{document}
\maketitle
\begin{abstract}
Retrieval-Augmented Generation (RAG) has become a powerful paradigm for enhancing large language models (LLMs) through external knowledge retrieval. Despite its widespread attention, existing academic research predominantly focuses on single-turn RAG, leaving a significant gap in addressing the complexities of multi-turn conversations found in real-world applications. To bridge this gap, we introduce CORAL, a large-scale benchmark designed to assess RAG systems in realistic multi-turn conversational settings. CORAL includes diverse information-seeking conversations automatically derived from Wikipedia and tackles key challenges such as open-domain coverage, knowledge intensity, free-form responses, and topic shifts. It supports three core tasks of conversational RAG: passage retrieval, response generation, and citation labeling. We propose a unified framework to standardize various conversational RAG methods and conduct a comprehensive evaluation of these methods on CORAL, demonstrating substantial opportunities for improving existing approaches. Our dataset and code are available at \url{https://github.com/Ariya12138/CORAL}.
\end{abstract}

\section{Introduction}\label{introduction} 
Retrieval-Augmented Generation (RAG) has emerged as a promising approach in question answering, leveraging large language models (LLMs) alongside external knowledge retrieval to enhance the quality and accuracy of generated responses~\citep{Lewis_nips20_rag,guu_realm,huang_rag_survey}. While RAG has gained traction both in academia and industry, a notable gap exists between how it is studied in academic settings and how it is implemented in real-world systems. Academic research predominantly focuses on single-turn interactions~\citep{tan_acl24_simplm,jin_acl24_bider,wang_richrag,dong2024toward,zhu_spring}, whereas most industrial RAG systems~\cite{chatgpt,claude,kimichat,doubao} are designed to handle multi-turn conversations. In practice, multi-turn conversation is the norm, where RAG systems must dynamically adapt to evolving context and user intent across multiple turns.

The shift from single-turn to multi-turn conversations introduces unique challenges for RAG. In multi-turn settings, systems must deal with redundant or irrelevant information from prior interactions and cope with abrupt topic shifts~\citep{ye_sigir24_convrag,topiocqa}. This complexity can degrade the retrieval and generation quality, especially as conversation histories grow, exacerbating the ``long context problem''~\citep{Ratner_ACL23_PCW,LongRoPE}. These issues highlight the need for dedicated research into multi-turn conversational RAG to address the realities of interactive, ongoing dialogue.

\begin{table*}
\centering
\setlength{\tabcolsep}{2.8pt} 
\footnotesize 
\scalebox{0.9}{
\begin{tabular}{lccccc}

\toprule
\textbf{Dataset} & \textbf{Open-domain} & \textbf{Knowledge-Intensive} & \textbf{Free-form Response} & \textbf{Topic Shift} & \textbf{Citation Labeling}\\ 
\midrule
CORAL (ours) & \cmark & \cmark & \cmark & \cmark & \cmark \\ 
\midrule
TopiOCQA~\citep{topiocqa} & \cmark & \cmark & \cmark & \cmark & \xmark \\ 
QReCC~\citep{Anantha_naacl21_cqr} & \cmark & \cmark & \cmark & \xmark & \xmark \\ 
Wizard of Wikipedia~\citep{wizard_of_wikipedia} & \cmark & \cmark & \cmark & \xmark & \xmark \\
CoQA~\citep{reddy2019coqa} & \xmark & \cmark & \cmark & \xmark & \xmark \\
OR-QuAC~\citep{or_quac} & \cmark & \cmark & \xmark & \xmark & \xmark \\

Doc2Dial~\citep{doc2dial} & \xmark & $\bigtriangleup$ & \cmark & \xmark & \xmark \\
TREC CAsT19~\citep{cast19} & \cmark & \cmark & \xmark & \xmark & \xmark \\
TREC CAsT20~\citep{cast20} & \cmark & \cmark & \cmark & \cmark & \xmark \\
TREC CAsT21~\citep{cast21} & \cmark & \cmark & \cmark & \cmark & \xmark \\
TREC CAsT22~\citep{cast22} & \cmark & \cmark & \cmark & \cmark & \xmark \\
\bottomrule
\end{tabular} }
\caption{Comparison of CORAL with other conversational search and conversational QA datasets. $\bigtriangleup$ indicates that only a portion of the dataset satisfies the property.}
\label{tab:dataset_comparison}
\end{table*}

However, progress in this area is severely hindered by the lack of a comprehensive benchmark designed to evaluate conversational RAG systems.
To align with the diverse and complex real-world applications of conversational RAG systems, we identify several critical features such a benchmark should satisfy: (1) open-domain coverage, allowing the system to handle questions from a wide range of topics; (2) knowledge-intensiveness, challenging systems to retrieve and generate responses that require deep, contextual knowledge; (3) free-form response generation, ensuring that models can produce detailed, contextually rich answers; (4) handling of topic shifts, evaluating the system’s ability to manage sudden changes in dialogue context without carrying over irrelevant information from previous turns; and (5) citation labeling, promoting transparency by requiring the system to cite the sources of the information it retrieves.


Unfortunately, no existing dataset satisfies all of these features.
Although there are datasets for related tasks, such as conversational search and question answering, they do not adequately address the unique challenges of benchmarking multi-turn conversational RAG systems that align with all of the above features. For instance, datasets in conversational search like TREC CAsT~\citep{cast19,cast20,cast21,cast22} primarily focus on retrieval tasks and lack the capacity to assess a system's generative abilities in producing free-form answers. Likewise, conversational QA datasets, such as QReCC~\citep{Anantha_naacl21_cqr} and TopiOCQA~\citep{topiocqa}, predominantly offer short, factual answers, falling short of reflecting the nuanced, long-form responses often required in practical applications. 
We summarize the limitations of existing related datasets in Table~\ref{tab:dataset_comparison}.
These limitations highlight the pressing need for a more comprehensive benchmark that meets the full spectrum of requirements for evaluating conversational RAG.

\begin{figure*}[ht]
  \centering
  \includegraphics[width=\linewidth]{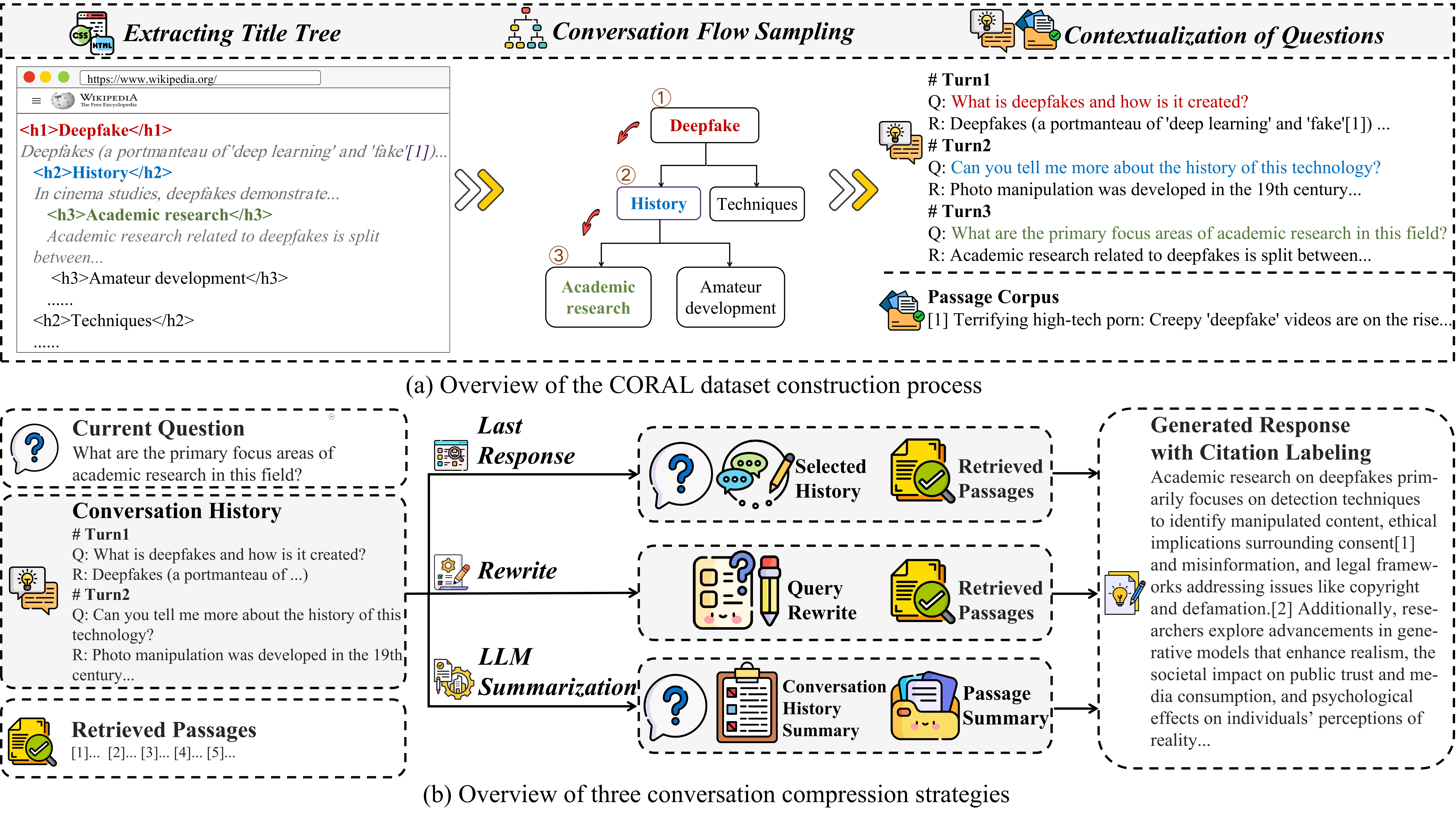}
  \caption{Part (a) is an overview of the CORAL dataset construction process. The red arrows show the sampled conversation flow, with numerical labels on the nodes indicating the round of the sampled conversation turns. The content under each sampled (sub)title serves as the conversational response in CORAL.
  Part (b) is the three conversation compression strategies in conversational RAG.}
  \label{fig:overview}
\end{figure*}

In this paper, we introduce a large-scale multi-turn \textbf{CO}nversational \textbf{R}etrieval-\textbf{A}ugmented Generation \textbf{L}anguage Benchmark (CORAL) that fulfills the above critical features to systematically evaluate and advance conversational RAG systems.
In general, CORAL is derived from English Wikipedia web pages, containing a total of 8,000 diverse information-seeking conversations. 
We propose a novel approach to automatically convert Wikipedia content into conversational formats, with each conversation generated through tailored sampling from either a single page or multiple related pages.
Specifically, as depicted in Figure~\ref{fig:overview}, we treat the (sub)titles of Wikipedia pages as the source of questions, using the corresponding human-written Wikipedia content serving as high-quality free-form responses. The content itself is originally well-cited and includes related passages for retrieval, making Wikipedia a particularly suitable source for constructing conversational RAG datasets.
To ensure coherent and diverse conversation flow, we design four sampling strategies based on the natural hierarchical properties of Wikipedia pages to first create the conversation flow. We then utilize powerful LLMs (e.g., GPT-4), to refine the original Wikipedia titles into well-formed conversational queries by incorporating contextual dependencies, such as co-reference and omission, resulting in the final conversation.

In CORAL, we evaluate conversational RAG systems across three essential tasks: (1) \textit{Conversational Passage Retrieval}, which assesses the system’s ability to retrieve the relevant information from a large document set based on multi-turn context; (2) \textit{Response Generation}, which tests the system’s capacity to generate accurate, contextually rich answers; and (3) \textit{Citation Labeling}, which ensures that the generated responses are transparent and grounded by requiring correct attribution of sources. These tasks are fundamental for measuring the core capabilities of conversational RAG systems in real-world, multi-turn settings.

Additionally, we present a unified framework that standardizes various conversational RAG baselines and conduct a comprehensive evaluation of them on the CORAL benchmark.
We find that the fine-tuned open-source LLM outperforms the commercial closed-source LLM in the retrieval stage, and shortening the input length to filter noise can not only maintain response quality but also improve citation labeling accuracy.



In summary, the contributions of our work are:

(1) We present an automatic and novel approach for constructing large-scale conversational RAG benchmarks from Wikipedia, leveraging its hierarchical structure and high-quality content to create diverse information-seeking conversations.

(2) The CORAL benchmark itself is unique in its comprehensive coverage of critical features, filling a significant gap in the evaluation of conversational RAG systems.

(3) We develop a unified framework for standardizing and evaluating various conversational RAG baselines, facilitating systematic comparison and advancement in this rapidly evolving field.





\section{Related Work}

\subsection{Retrieval-Augmented Generation}
Existing RAG studies primarily focus on the optimization of individual components.
The rewriter~\cite{li_sigir24,wang_emnlp23_query2doc,baek_arxiv} module interprets and reconstructs user queries to align them more effectively with the search process. The reranker~\cite{ma_acl23,dong_dparag,xu_www24_genpt} module independently adjusts the ordering of retrieved documents based on their relevance. The post-retrieval processing~\cite{xu_recomp,yang_emnlp23_prca,wang_filco,jiang_acl24_longllmlingua,jin_acl24_bider} module then reduces the volume of these documents, stripping away non-essential information to focus solely on the content critical for generating precise responses.

While these works have significantly advanced single-turn RAG systems, gaps remain in multi-turn conversational settings~\cite{ye_sigir24_convrag}. We address this by proposing a new benchmark and framework specifically for conversational RAG.

\subsection{Conversational Search}
Conversational search enables users to interact with retrieval systems through multi-turn dialogues~\citep{mo2024survey}. Two main approaches are conversational query rewriting (CQR) and conversational dense retrieval (CDR). CQR transforms context-dependent queries into fully rewritten versions for ad-hoc retrieval, focusing on selecting relevant tokens from the conversation history~\citep{Voskarides_sigir20_cqr,Kumar_emnlp20_cqr,lin_tois21_cqr} or using LLMs to generate rewrites~\citep{lin_arxiv_20,yu_sigir20_cqr,Vakulenko_wsdm21_cqr,wu_emnlp22_conqrr}. CDR jointly encodes conversation history and the current query for end-to-end dense retrieval~\citep{yu_sigir21_cdr,mao_arxiv_chatretriever}.

Challenges like limited training data are addressed through data augmentation~\citep{lin_emnlp21_cdr, mao_emnlp22_convtrans, dai_icml22_impainting, jin_emnlp23_instructor, chen_acl24_cqr, mo_www24_convsdg}, and context denoising~\citep{mao_sigir22_coted,mo_kdd23_cqr,mao_www23_lecore,mo_acl23_cqr} improves retrieval by filtering irrelevant conversation history. However, a benchmark is still needed for evaluating response generation and citation labeling.

\begin{figure*}[t]
  \centering
  \includegraphics[width=\linewidth]{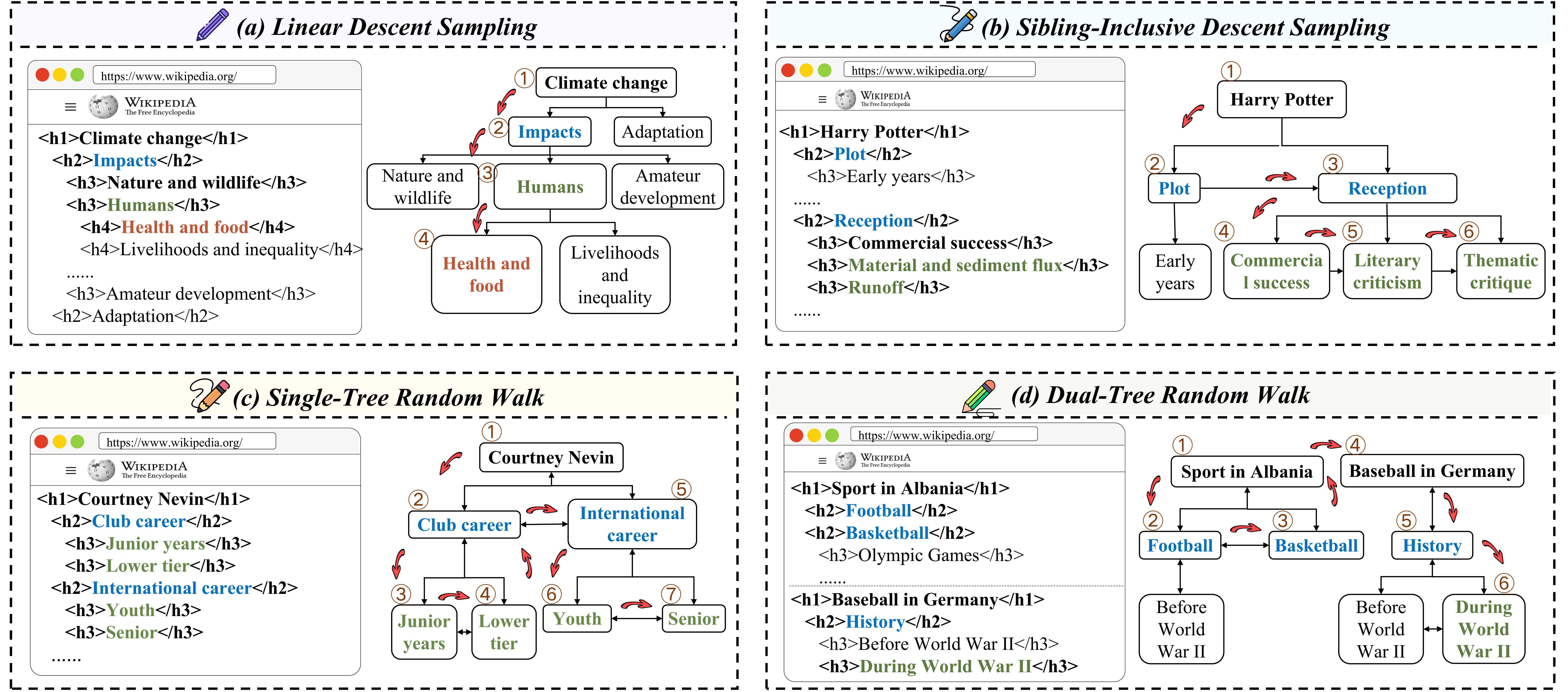}
  \caption{Illustration of the four sampling strategies. The red arrows show the sampled conversation flow, with numerical labels on the nodes indicating the round of the sampled conversation turns.}
  \label{fig:title_tree}
\end{figure*}

\section{CORAL}


\subsection{Data Source}
\label{sec:data_source}
We choose Wikipedia as our data source for the following reasons, which align with the critical features in Table~\ref{tab:dataset_comparison}.
(1) Wikipedia pages are well-structured and enriched by global volunteers, covering a broad range of topics;
(2) The logically interconnected titles provide a strong foundation basis for generating diverse queries, with each representing a distinct intent.
(3) The human-authored content under each title includes references that not only allow for free-form responses with precise citation labeling but also serve as the golden retrieval evidence for their respective titles.


However, the content may include noisy text, and reference pages are often too long for effective retrieval. 
We follow previous work~\cite{qian_webbrain} to clean the Wikipedia pages. Specifically,
for content, we remove Wikipedia templates, special symbols, and other invalid text. 
For references, we first split the reference pages into smaller passages. Then, we exclude passages shorter than 16 words or with a non-English token ratio exceeding 0.3, and then calculate term recall to identify suitable passages. After these refinements, we generate a clean set of 20,000 high-quality pages for subsequent conversations.

\subsection{The CORAL Dataset Construction}
We transform one or more related Wikipedia web pages into information-seeking conversations through a three-stage approach.

\subsubsection{Extracting Title Trees}
First, we extract all subheadings (i.e., titles) from the raw HTML of the Wikipedia pages. These subheadings create a natural hierarchy for the content, enabling us to construct a title tree, where the page title (H1-level heading) serves as the root. Subsequent headings (e.g., H2 to H6) divide the content into progressively detailed sections, with each level corresponding to a node's depth in the tree. The directional links between nodes will dictate the flow of the generated conversations.
Besides, to enhance the complexity and diversity of conversations, we also adjust the depth, breadth, multi-subtopic exploration, and topic shifts during the construction of these title trees.

\subsubsection{Conversation Flow Sampling}\label{sec:conv_flow}
To generate coherent and diverse conversations, we implement the following four sampling strategies based on the extracted title trees:

(1) \textit{Linear Descent Sampling (LDS)}: This strategy begins at the root node and permits movement only from parent nodes to their child nodes. LDS serves as the most basic sampling path, emulating the progressive logic typical of real conversational information-seeking scenarios. As illustrated in Figure~\ref{fig:title_tree}(a), the title tree starts with the overall theme of climate change and progressively narrows down to specific impacts associated with this global issue. Following the red arrow, the focus shifts to the human aspects, particularly examining how climate change affects human health and food security. This structure exemplifies a gradual deepening of the query intent as the conversation unfolds.

(2) \textit{Sibling-Inclusive Descent Sampling (SIDS)}: This strategy builds on LDS by introducing directional links between sibling nodes. This feature is essential because conversational processes often encompass both in-depth and parallel explorations of related subtopics.
As shown in Figure~\ref{fig:title_tree}(b), when discussing the reception of Harry Potter, the subsequent three rounds of dialogue analyze it from three distinct perspectives: commercial success, literary criticism, and thematic critique. This enhancement enriches the breadth of discussions within the conversation structure.

(3) \textit{Single-Tree Random Walk (STRW)}: This strategy further enhances SIDS by incorporating interconnections among sibling nodes as well as between parent and child nodes. Essentially, it forms a directed graph with bidirectional edges. As illustrated in Figure~\ref{fig:title_tree}
(c), after an in-depth exploration of Courtney Nevin's club career, the focus shifts to her international career.

(4) \textit{Dual-Tree Random Walk (DTRW)}: It mimics the topic shifts that occur in real conversational scenarios, allowing for greater flexibility. It enables transitions between two different but somewhat related trees, which are retrieved using the root title as a query and employing the BM25 algorithm. As illustrated in Figure~\ref{fig:title_tree}(d), the conversation shifts from sports in Albania to baseball in Germany.

\subsubsection{Contextualization of Questions}
\label{question_generation}
As introduced in Section~\ref{sec:data_source}, we treat the subtitles as the sources of questions, with their corresponding contents serving as the responses. In this final stage, we contextualize the keyword subtitles into conversational questions to enhance the realism of the conversation.

Specifically, for each turn, we first create a keyword chain that includes the current node and all its ancestor nodes. This keyword chain, along with the response of the current node, is then used to prompt GPT-4\footnote{gpt-4-turbo-2024-04-09 from https://openai.com/api\label{gpt-4}} to rewrite the original keyword title into a natural language question. We then continue to prompt GPT-4 to further contextualize the question into a conversational format by incorporating linguistic phenomena such as ellipses, references, and omissions~\citep{cast19}, which are prevalent in real conversational scenarios.
The prompt details are provided in Appendix~\ref{prompts_of_contextualization_of_questions}.

\begin{table*}[htbp]
    \centering
    \setlength{\tabcolsep}{12pt}
    \small 
    \begin{tabular}{@{}ccccccc@{}}
        \toprule
        Category & Method & MRR & MAP & NDCG@3 & Recall@20 & Recall@100 \\
        \midrule
        
        \multirow{4}{*}{CDR Models} &Conv-ANCE-Q & 19.8 & 28.6 & 20.5 & 39.1 & 51.0 \\
        &KD-ANCE-Q & 22.6 & 33.1 & 24.5 & 38.5 & 48.0 \\
        &Conv-ANCE-C & 20.5 & 29.6 & 21.1 & 39.8 & 53.4 \\
        &KD-ANCE-C & \textbf{23.2} & \textbf{33.6} & 24.9 & \textbf{40.3} &\textbf{49.6} \\
        
        \midrule
        \multirow{3}{*}{CQR Models} &LLM4CS (GPT-3.5) & 21.2 & 31.1 & 23.0 & 35.5 &44.4 \\
        &Qwen2.5-1.5B & 16.3 & 23.8 & 17.2 & 31.0 & 39.2 \\
        &Qwen2.5-1.5B-SFT & 23.1 & \textbf{33.6} & \textbf{25.1} & 39.4 &48.6 \\
        \bottomrule
    \end{tabular}
    \caption{Retrieval performance comparisons. The best performance is \textbf{bold}.
    \textit{Conv-ANCE-Q} denotes the Conv-ANCE is trained on the QReCC dataset and \textit{Conv-ANCE-C} denotes the Conv-ANCE is trained on CORAL training dataset.}
    \label{conversational_search_performances}
\end{table*}

\subsection{The Final Dataset Format and Statistics}
The key statistics of CORAL are summarized in Table~\ref{statistics_four_conversation_structures}. Our dataset consists of 8,000 conversations with the four types introduced in Section~\ref{sec:conv_flow}. These 8,000 conversations are evenly distributed across four distinct structural types, with each type containing 2,000 conversations. Specifically, the LDS conversation type includes 3 to 6 turns per conversation. For the remaining types—SIDS, STRW, and DTRW—each category consists of 1,600 sets of conversations with 6 to 10 turns, along with an additional 400 sets featuring 11 to 20 turns per conversation. The design of the longer conversation intends to simulate real-world challenges encountered in conversational scenarios, such as redundant information and the long context problem. 

Our final dataset format is as follows: A conversation $\mathbf{C}=\{ (q_{i},r_{i})\}_{i=1}^{n}$ comprised of $n$ turns. $q_{i}$ is a contextualized query of the $i$-th turn generated in Section~\ref{question_generation}, and $r_{i}$ is the $i$-th turn golden response, which is the cleaned plain text under the corresponding (sub-)title in the HTML.
The supporting web pages for $r_{i}$, listed in the HTML Reference Section, can be processed as described in Section~\ref{sec:data_source} to serve as the golden relevant passages $P_i^+ = \left\{ p_{i,1}, p_{i,2}, \ldots \right\}$.
On average, each conversation turn has 3.17 related passages, and the average golden response length is 255 tokens. Finally, we obtain a passage corpus $\mathcal{P} $, which contains 200K passages from all the golden references $P_i^+$.


\subsection{Evaluation Tasks}
CORAL mainly supports three fundamental conversational RAG tasks:

(1) \textit{Conversational Passage Retrieval}:
This task evaluates a system's capability to extract relevant information from extensive document collections, considering the context of multi-turn conversations. Formally, given the $k$-th question $q_{k}$ and the corresponding conversation history
$H_{k} = \{ q_{i},r_{i}\}_{i=1}^{k-1}$, where $q
_{i}$ and $r_{i}$ respectively denote the question and response of the $i$-th turn, the retriever $\mathbb{R} $ aims to retrieve the relevant passages
$P_{k} $
from the passage corpus $\mathcal{P} $. 
We use
MRR, MAP, NDCG@3, Recall@20 and
Recall@100 as retrieval evaluation metrics.

(2) \textit{Response Generation}: This task challenges the system's ability to produce accurate, detailed, and contextually appropriate answers.
Given the $k$-th question $q_{k}$, the corresponding conversation history $ H_{k} $, and the relevant passages  $P_{k}$, the generator $\mathbb{G} $ needs to generate an informative response to answer the question. We use rule-based metrics BLEU-1~\cite{bleu}, and ROUGE-L~\cite{lin2004rouge} to evaluate the response quality compared with $r_{k}$. Given the lengthier responses in our benchmark, we additionally utilize the model-based evaluation method proposed in RichRAG~\citep{wang_richrag}.

(3) \textit{Citation Labeling}: This task evaluates the method's ability to accurately attribute information sources within the generated responses.
Following ALCE~\citep{gao_alce_emnlp23}, the generated response $r_{k}$ consists of $n$ statements $s_1, s_2,...,s_n$. Each statement $s_i$ cites a list of passages $C_i = \{ c_{i,1},c_{i,2},...\} $, where $c_{i,j} \in P_k$. We adopt Citation Recall and Citation Precision defined in ALCE~\citep{gao_alce_emnlp23} to evaluate the accuracy of citation labeling.




\begin{table*}[tbp]
    \centering
    \setlength{\tabcolsep}{11pt}\renewcommand{\arraystretch}{0.9}
    \setlength{\tabcolsep}{4pt}
    \small 
    \begin{tabular}{@{}p{2.8cm}p{1.5cm}lcccc@{}}
        \toprule
        \multirow{2}{*}{Category} & \multirow{2}{*}{\textbf{\# Tokens}} &\multirow{2}{*}{Model} & \multicolumn{2}{c}{Generation} & \multicolumn{2}{c}{Citation Labeling} \\
        \cmidrule(lr){4-5}
        \cmidrule(lr){6-7}
        & & &BLEU-1 & ROUGE-L & Citation Recall & Citation Precision \\
        \midrule

        \multirow{6}{2.8cm}{Raw Context} & \multirow{6}{*}{2226} &Qwen2.5-7B &  22.2  & 13.1&3.1&18.1 \\
        &&Mistral-7B & 18.1 & 12.4 &2.4 &4.8\\
        &&Llama-3.1-8B & 21.5 & 12.9 & 0.9 & 2.1\\
        \cmidrule(lr){3-7}
        &&Qwen2.5-7B-SFT & 18.3 &  18.5 & 6.6& 16.8\\  
        &&Mistral-7B-SFT & 23.7  & \textbf{20.1} & 4.6 & 11.1\\
        &&Llama-3.1-8B-SFT & 24.2 & 19.7 & 4.2 & 9.8\\

        \midrule
        
        \multirow{6}{*}{Last Response} &\multirow{6}{*}{1474} &Qwen2.5-7B & 20.9 & 12.8 & 3.5 & 20.8 \\
        &&Mistral-7B & 18.1  & 12.3 & 2.7 & 4.5\\
        &&Llama-3.1-8B & 20.4 & 12.7 & 1.3 & 3.1 \\
        \cmidrule(lr){3-7}
        &&Qwen2.5-7B-SFT & 23.9  & 16.5 & 10.4 & 24.8\\  
        &&Mistral-7B-SFT & 21.8 & 18.5 & 5.0 & 12.4\\
        &&Llama-3.1-8B-SFT & 26.1 & 18.1 & 3.5 & 8.7\\

        \midrule
        \multirow{6}{*}{Rewrite} &\multirow{6}{*}{1236} 
        &Qwen2.5-7B & 21.1  & 12.8 & 2.4& 9.4 \\
        &&Mistral-7B & 18.8  & 12.3 & 2.5 & 3.8 \\
        &&Llama-3.1-8B & 18.8 & 12.4 & 1.7 & 3.1 \\
        \cmidrule(lr){3-7}
        &&Qwen2.5-7B-SFT & 18.9 & 16.4 & 7.4 & 16.8\\
        &&Mistral-7B-SFT & 24.8 & 18.5 & 5.9 & 14.8\\
        &&Llama-3.1-8B-SFT & \textbf{26.3} & 18.2 & 4.7 & 11.2\\

        \midrule
        \multirow{6}{4cm}{LLM Summarization} & \multirow{6}{*}{478} 
        &Qwen2.5-7B & 21.0 & 12.7 & 2.9 & 13.0 \\
        &&Mistral-7B & 19.5 & 12.3 & 5.6 & 6.7 \\
        &&Llama-3.1-8B & 19.1 & 12.8 & 4.1 & 7.1 \\

        \cmidrule(lr){3-7}
        &&Qwen2.5-7B-SFT & 23.5&16.8& \textbf{14.1} & \textbf{31.1}\\
        &&Mistral-7B-SFT & 16.9 & 17.1 & 8.3 & 19.8 \\
        &&Llama-3.1-8B-SFT & 18.7 & 16.5  & 4.5 & 10.7 \\
        
        \bottomrule
    \end{tabular}
    \caption{The comparison of different LLMs on response generation and citation labeling. \textit{\# Tokens} denotes the number of input tokens. }
    \label{generation_performances}
\end{table*}


\section{Conversational RAG Framework}
\label{sec:conversation_compression}
A conversational RAG system typically comprises a retriever and a generator to handle the current user query $q_k$, the conversation history $H_k$, and the retrieved passages $P_k$. As the conversation progresses, both the growing conversation history and the noisy retrieved passages can negatively impact the system's efficiency and effectiveness, making it harder to generate accurate responses. To solve the problem, we propose a simple compression framework to efficiently manage these inputs.
Specifically, we introduce a conversation compression function $f$ to compress the conversation, and then use the compressed contents as the real inputs of retrievers and LLM generators. In addition to conversation compression, we also apply post-retrieval results compression. Following existing approaches~\cite{xu_recomp}, we simply take LLMs as the compression function $f_{p}$, leaving the exploration of more compression methods in future work.

Formally, suppose $f(H_k)$ is the compressed conversation context, $P_k=\mathbb{R} (f(H_k), q_k)$ is list of passages retrieved by querying $f(H_k)$ with $q_k$, and $f_{p}(P_k)$ is the compressed results of the retrieval,
the final generation task can be formulated as:
$\mathbb{G}(q_k, f(H_k), f_{p}(P_k)) $. The prompt for feeding $q_k, f(H_k), f_{p}(P_k)$ into the generator can be found in Appendix~\ref{prompts_of_generating_responses}.
Various existing conversational RAG methods can be unified into our framework. In this work, we mainly investigate the following three methods for the conversation compression:

\paragraph{Last Response Strategy} 
For the conversation history, 
we heuristically select all previous conversational questions $\{q_{i}\}_{1}^{k-1}$ and the last turn's response $r_{k-1} $ in the conversation history:
\begin{eqnarray}
    f^{\text{LR}}(H_k)=\{q_{i}\}_{1}^{k-1},r_{k-1}.
\end{eqnarray}

\paragraph{Rewrite Strategy}
We adopt a conversational query rewriting model Rewrite() to transform the original query along with the conversation history into a standalone question
rewrite $\hat{q_k}$:
\begin{eqnarray}
    f_{c}^{\text{RW}}(H_k) =\hat{q_k} = \text{Rewrite}\left(q_k;H_k \right).    
\end{eqnarray}
In this strategy, $P_k=\mathbb{R} (\hat{q_k})$ is list of passages retrieved by querying $\hat{q_k}$, and $f_{p}(P_k)$ is the compressed results of the retrieval,
the final generation task can be formulated as:
$\mathbb{G}(\hat{q_k}, f_{p}(P_k)) $.

\paragraph{LLM Summarization Strategy}
Inspired by RECOMP~\citep{xu_recomp}, we use LLMs to generate abstractive summary of the conversation history:
\begin{eqnarray}
    f_{c}^{\text{SUM}}(H_k) = \text{LLM}(H_{k}).
\end{eqnarray}
The prompt is shown in Appendix~\ref{prompts_of_LLM_summarization_strategy}.



\section{Experiments}

In this section, we discuss the performance of conversational RAG on our benchmark, and provide a comprehensive analysis for each stage.

\subsection{Evaluating Retrieval Performance}

We concentrate on two main approaches in conversational search: conversational dense retrieval (CDR) and conversational query rewriting (CQR). For CDR, we use KD-ANCE and Conv-ANCE with ANCE as the base retriever. KD-ANCE~\citep{yu_sigir21_cdr} trains the session encoder by mimicking golden query embeddings, while Conv-ANCE~\citep{Karpukhin_dpr_emnlp20,lin_emnlp21_cdr} uses contrastive learning to train the session encoder, drawing it closer to relevant passages and further from irrelevant ones. For CQR, we utilize the LLM4CS~\citep{mao_emnlp23_llm4cs}, which incorporates GPT-3.5, and an open-source LLM for generating query rewrites respectively to enable a comparative analysis. Table~\ref{conversational_search_performances} provides a detailed comparison between these two categories. We have the following observations:

\begin{figure*}[!t]
  \centering
  \includegraphics[width=\linewidth]{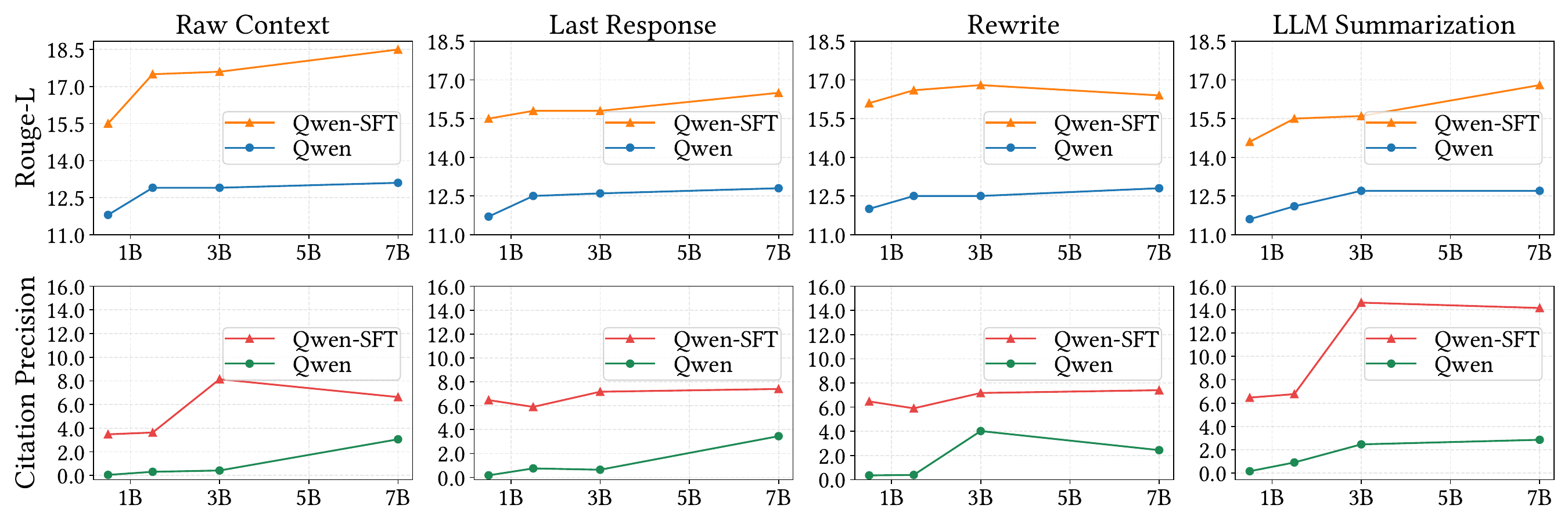}
  \caption{The scaling analysis of generation and citation labeling performance.}
  \label{fig:scaling}
\end{figure*}

\begin{figure*}[!t]
  \centering
  \includegraphics[width=\linewidth]{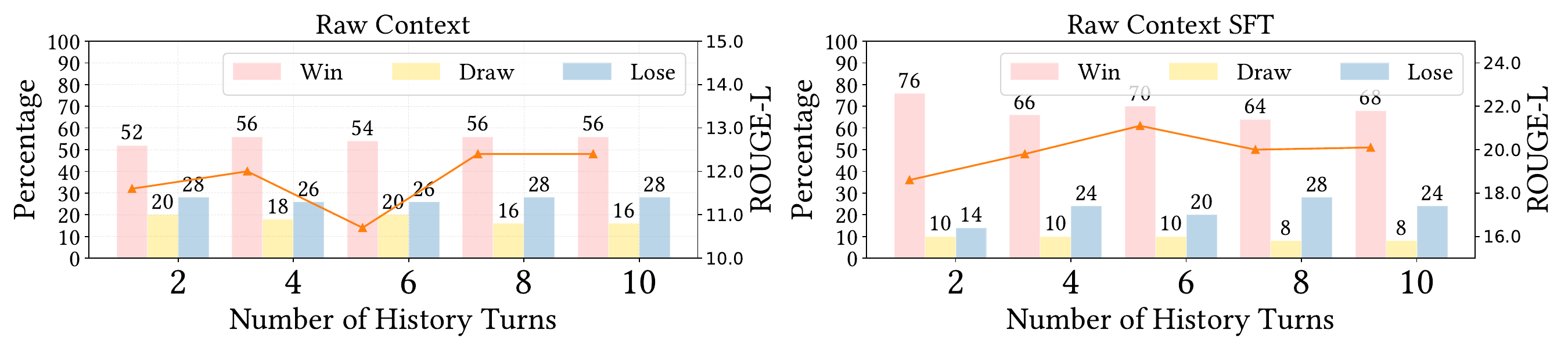}
  \caption{Generation results of different conversation history length. The curve in the figure represents the ROUGE-L score. The histogram shows the results of GPT-4 scores comparing model-generated responses with golden responses. \textit{Win} indicates cases where model-generated responses outperform golden responses, \textit{Draw} indicates cases where the two responses are considered equally good, and \textit{Lose} indicates cases where the golden responses are considered better. The y-axis on the left represents the proportion of cases in the total number of cases.}
  \label{fig:irrelevant_conversation_information}
\end{figure*}

\begin{figure}[h]
  \centering
  \includegraphics[width=\linewidth]{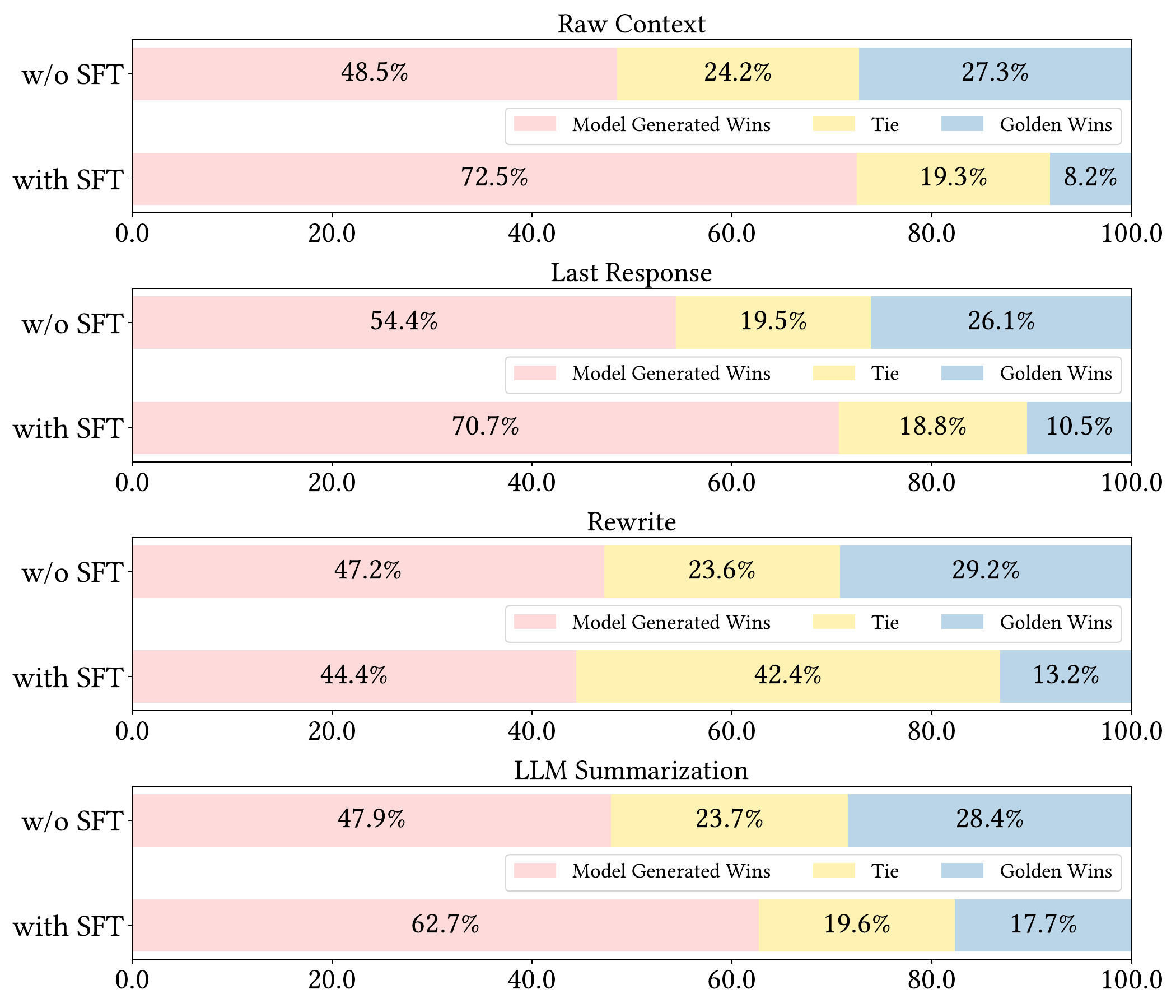}
  \caption{The GPT-4 evaluation score.}
  \label{fig:gpt_evaluation}
\end{figure}

(1) The performances of the CDR and CQR models are fairly comparable. Notably, the Qwen2.5-1.5B-SFT shows a strong competitive edge, not only surpasses the Qwen2.5-1.5B but also outperforms the advanced closed-source LLM GPT-3.5 across all evaluated metrics.

(2) KD-ANCE in the CDR category shows better results compared to Conv-ANCE. This may be attributed to the training methodologies: KD-ANCE possibly leverages golden rewrite data more effectively than Conv-ANCE, which uses in-batch negatives that may not be sufficiently challenging for optimal learning.





\subsection{Evaluation Response Generation with Citation Labeling}
We compare the raw context baselines with another three conversation compression strategies introduced in Section~\ref{sec:conversation_compression}, selecting Qwen2.5, Mistral, and Llama as generators. We prompt the generator to generate the response along with the citations in the response.
The generation and citation labeling performance is shown in Table~\ref{generation_performances}, and the GPT-4 score is shown in Figure~\ref{fig:gpt_evaluation}. We find that:

(1) By examining four methods of modeling conversation history, we observe an interesting trend: as the input is progressively condensed (from 2226 input tokens in the raw context to merely 478 input tokens in the LLM Summarization), 
the decrease in performance is surprisingly minimal, and in terms of citation labeling, there is even an observed improvement. This suggests that some content within the dialogue history is irrelevant or redundant and can be removed without negatively impacting the model's performance.


(2) Among three conversation compression strategies, the Rewrite with SFT exhibits superior performance, which could be attributed to the model's enhanced capability to learn from the simplified question-answer pattern. Intriguingly, although the LLM Summarization strategy demonstrates weaker performance in response generation, it significantly enhances citation labeling. A possible explanation is that the summarization process effectively filters noise, thereby optimizing the content for generating more reliable responses.

\subsection{Scaling Analysis on Model Parameters}
We scale the generator's parameters from 500M to 7B, as shown in Figure~\ref{fig:scaling}. We find that: 

(1) There is a pronounced improvement in generation as parameters increase from 500M to 1.5B, evidenced by a significant rise in ROUGE-L scores. 
However, beyond 3B parameters, the performance gains plateau, indicating diminishing returns with additional parameter scaling.

(2) Performance in citation labeling improves markedly as the parameter count extends from 3B to 7B. This suggests that a larger model capacity is beneficial for tasks that require extensive knowledge, such as accurate citation usage.

\subsection{Quantitative Analysis on History Turns}
To analyze the impact of conversation history length, we randomly select 50 conversations and vary the number of previous dialogue turns provided to the generator. This can be represented as $r_{k}^m=\mathbb{G}  \left ( q_{k};H_{k}^m;P_{k} \right ) $, where $k=12$, $H_{k}^m = \{ q_{i},r_{i}\}_{k-m}^{k-1} $, and $m \in \{2, 4, 6, 8, 10\}$. Results are shown in Figure~\ref{fig:irrelevant_conversation_information}. We find that:

(1) After fine-tuning, the performance improves significantly, especially when using four history turns, resulting in a notable 55\% improvement in the ROUGE-L. This demonstrates the effectiveness of SFT in modeling history.

(2) Before fine-tuning, response quality decreases with six turns of history compared to four, possibly due to the redundant information introduced by the longer history. However, after fine-tuning, response quality improves with six turns but declines with eight, suggesting a trade-off between richer information enriched by longer context and irrelevant information introduced by conversation history. These findings validate the challenges previously discussed in Section~\ref{introduction}.


\section{Conclusion}
In this paper, we present an automatic approach using LLMs to construct large-scale, information-seeking conversations from Wikipedia pages. The resulting benchmark, CORAL, supports three fundamental tasks for evaluating conversational RAG systems. Additionally, we propose a unified framework to standardize various conversational RAG methods and conduct a comprehensive evaluation of these methods on CORAL. We envision CORAL as a valuable resource for advancing research in conversational RAG, fostering innovation, and improving real-world applications.


\section*{Limitations}

Our work presents a conversational RAG benchmark named CORAL, which fills a notable void in assessing conversational RAG methods. In this benchmark, we examine the effects of compressing conversational history on answer generation, paving the way for future research in conversational RAG. However, since CORAL is built upon Wikipedia and existing LLMs are typically trained on corpora like Wikipedia and CommonCrawl, using these LLMs as generators could lead to contamination in the conversational RAG process due to the overlap in their training data. Additionally, the three conversation compression strategies employed in CORAL are somewhat basic, focusing solely on reducing the length of inputs rather than modeling the conversation history in a granular manner. Additionally, the use of the LLM Summarization strategy for compressing both conversation history and retrieved passages, while leveraging advanced models such as GPT-4, could lead to considerable expenses.

\bibliography{custom}

\appendix

\clearpage
\section*{Appendix}

\begin{table*}[h]
\centering
\relsize{-1}
\begin{tabular}{lcccccccc}
\toprule

& \multicolumn{2}{c}{LDS}& \multicolumn{2}{c}{SIDS} &  \multicolumn{2}{c}{STRW} & \multicolumn{2}{c}{DTRW}\\
\cmidrule(lr){2-3}
\cmidrule(lr){4-5}
\cmidrule(lr){6-7}
\cmidrule(lr){8-9}

 & Train &Test & Train &Test & Train &Test &Train &Test\\

 \midrule
\# Conversation & 1800 & 200 & 1800 & 200 & 1800 & 200 & 1800 & 200 \\
\midrule
\# Turns & 5934 & 651 & 16082 & 1727 & 18165 & 1949 & 19411 & 2153\\
\midrule
\# Turns / Conversation & 3.30 & 3.26 & 8.93 & 8.64 & 10.09 & 9.75 & 10.78 & 10.77\\
\midrule
\# Tokens / Question &13.70 &13.89 & 12.62 & 12.64 & 12.72 & 12.88 & 14.15 &14.75\\
\midrule
\# Tokens / Response & 233.81 & 147.16 & 242.54 & 155.54 & 243.34 & 191.60 & 300.47 & 259.72\\
\midrule
\# Positive passages/ Turn & 3.25 & 2.03 & 2.64 & 1.73 & 3.01 & 2.12 & 3.98 & 3.50\\
\bottomrule
\end{tabular}
\caption{Data statistics of four different conversation structures.}
\label{statistics_four_conversation_structures}
\end{table*}

\section{Prompts of the Contextualization of Questions}
\label{prompts_of_contextualization_of_questions}

When contextualizing questions, two steps need prompt. Firstly, we transform the node into the complete question. Secondly, we convert the complete questions into conversational questions.
Table~\ref{prompt_of_generating_question} illustrates the prompt for generating a complete question.
Table~\ref{prompt_of_generating_conversational_question} demonstrates the prompt for creating conversational questions. Following LLM4CS~\citep{mao_emnlp23_llm4cs}, the prompt consists of three components: Instruction, Demonstration, and Input.
The red section is designated for LDS prompting, the blue section for SIDS and STRW prompting, the green section for DTRW prompting, and the orange section for LDS, SIDS, and STRW prompting.

\section{Prompts of Generating Responses with Citation Labeling}
\label{prompts_of_generating_responses}
Table~\ref{prompt_of_generating_question} provides the prompt template for generating a response with citation labeling. The red part is for Raw Context and the Last Response strategy prompting. 
The blue part is for Rewrite Strategy prompting. 
The green part is for LLM Summarization Strategy prompting. The orange part is for Raw Context, the Last Response Strategy, and the LLM Summarization Strategy prompting.

\section{Prompts of LLM Summarization Strategy}
\label{prompts_of_LLM_summarization_strategy}
Table~\ref{prompt_of_summarizing_conversation_history} provides a general illustration of the prompt of generating a summary of conversation history. 

\section{More Detailed Experimental Setting}

\subsection{Conversational Search Baselines}
The conversational search baseline models are chosen for their prevalence and effectiveness in the field. 
We focus on two primary approaches: conversational dense retrieval (CDR) and conversational query rewriting (CQR). 
For CDR, we adopt KD-ANCE and Conv-ANCE, where ANCE is a base ad-hoc retriever. Following ~\citet{yu_sigir21_cdr}, KD-ANCE uses an ad hoc query encoder as the teacher model, training the student session encoder to imitate the embeddings derived from the golden queries. Meanwhile, according to the methodology outlined by~\citep{Karpukhin_dpr_emnlp20,lin_emnlp21_cdr}, Conv-ANCE is designed to implement the classical ranking loss function. This function strives to minimize the distance between the session and its relevant passages while maximizing the separation from irrelevant ones.  Dense retrieval is conducted using Faiss~\cite{faiss}.
For CQR, we choose LLM4CS~\cite{mao_emnlp23_llm4cs}, employing the proprietary commercial model GPT-3.5 to generate rewrites. Additionally, we choose an open-source LLM to generate rewrites as well, allowing for a comparative analysis.

\subsection{Generation with Citation Labeling}
We compare the raw context baselines with another three conversation compression strategies introduced in Section~\ref{sec:conversation_compression}. We choose Qwen2.5-7B-Instruct~\citep{qwen2,qwen2.5}, 
Mistral-7B-Instruct~\citep{mistral}, and Llama-3.1-8B-Instruct~\citep{llama3} as the generator. For the scaling analysis, we use the Qwen2.5-Instruct series, specifically the 0.5B, 1.5B, 3B, and 7B models, as our generators for detailed examination. 
During the training process, we utilize the LLaMA-Factory~\citep{zheng2024llamafactory} framework, running on two A800 GPUs. The training parameters are set as follows: we employ a learning rate of 1.0e-5. The batch size is maintained at 1, and the maximum token length for training instances is set to 4096. Because of the lack of training data of the LLM Summarization category, we use the checkpoint of Raw Context.

During the inference process, we leverage the vLLM~\citep{kwon2023efficient} framework to accelerate inference.  The maximum input length is set to 32,000, top\_p is set to 0.9, and temperature is set to 1.

\subsection{More detailed Scaling Analysis}
Table~\ref{tab: comparison_of_different_size} provides detailed results of the generation quality and citation labeling accuracy.

\section{Dataset Format}
Table~\ref{example_of_coral} provides an example of CORAL. Our dataset CORAL has information-seeking questions, free-form responses with citation labeling, golden rewrites, and corresponding golden retrieval passage ID.

\begin{table*}[hb]
    \centering
    \relsize{-1} 
    \begin{tabular}{@{}llcccc@{}}
        \toprule
        \multirow{2}{*}{Category} &\multirow{2}{*}{\textbf{Model}} & \multicolumn{2}{c}{Generation} & \multicolumn{2}{c}{Citation Labeling} \\
        \cmidrule(lr){3-4}
        \cmidrule(lr){5-6}
        & & BLEU-1 & ROUGE-L & Citation Recall & Citation Precision \\
        \midrule

        \multirow{8}{*}{Raw Context} 
        &Qwen2.5-0.5B & 16.4  & 11.8 &0.1 &0.2 \\
        &Qwen2.5-1.5B & 20.8 & 12.9 & 0.3 & 1.2 \\
        &Qwen2.5-3B &  21.4  & 12.9 & 0.4 & 1.8\\
        &Qwen2.5-7B &  22.2  & 13.1 &3.1 &18.1\\
        \cmidrule(lr){2-6}
        &Qwen2.5-0.5B-SFT & 13.0  & 15.5 & 3.5 & 10.2\\
        &Qwen2.5-1.5B-SFT & 20.9  &17.5 &3.6 &10.0\\
        &Qwen2.5-3B-SFT & \textbf{25.8}  & 17.6 & 8.1 & 20.7 \\
        &Qwen2.5-7B-SFT & 18.3  & \textbf{18.5} & 6.6 &16.8\\  
        \midrule

        \multirow{8}{*}{Last Response} 
        &Qwen2.5-0.5B & 15.6 & 11.7 & 0.2 & 0.5 \\
        &Qwen2.5-1.5B & 19.3 & 12.5 & 0.8 & 2.9 \\
        &Qwen2.5-3B & 21.1 & 12.6 & 0.6 & 3.4  \\
        &Qwen2.5-7B & 20.9  & 12.8 & 3.5 & 20.8 \\
        \cmidrule(lr){2-6}
        &Qwen2.5-0.5B-SFT & 19.8 & 15.5 & 6.5 & 18.0\\
        &Qwen2.5-1.5B-SFT & 19.4 & 15.8 & 6.7 & 17.7\\
        &Qwen2.5-3B-SFT & 21.8 & 15.8 & 7.4 & 17.1\\
        &Qwen2.5-7B-SFT & 22.1 & 16.5 & 10.4 & 24.8\\

        \midrule
        \multirow{8}{*}{Rewrite} 
        &Qwen2.5-0.5B & 17.3 & 12.0 & 0.4 & 0.8 \\
        &Qwen2.5-1.5B & 19.9 & 12.5& 0.4 & 1.2 \\
        &Qwen2.5-3B & 20.8 & 12.5 & 4.0 & 14.9 \\
        &Qwen2.5-7B & 21.1 & 12.8 & 2.4& 9.5 \\
        \cmidrule(lr){2-6}
        &Qwen2.5-0.5B-SFT & 21.4 & 16.1 & 6.5 & 16.5\\
        &Qwen2.5-1.5B-SFT & 21.7 & 16.6 & 5.9 & 14.9\\
        &Qwen2.5-3B-SFT & 23.3 & 16.8 & 7.2 & 16.5\\
        &Qwen2.5-7B-SFT & 18.9 & 16.4 & 7.4 & 19.8\\

        \midrule
        \multirow{8}{*}{LLM Summarization} 
        &Qwen2.5-0.5B & 13.2&11.6 & 0.2 & 0.4 \\
        &Qwen2.5-1.5B & 15.0 & 12.1 &0.9 & 3.0\\
        &Qwen2.5-3B & 20.2 & 12.7 & 2.5 &10.6\\
        &Qwen2.5-7B & 21.0 & 12.7 & 2.9 & 13.0\\
        \cmidrule(lr){2-6}
        &Qwen2.5-0.5B-SFT & 21.4 & 14.6 & 6.5 & 17.4\\
        &Qwen2.5-1.5B-SFT & 23.0 & 15.5 & 6.8 & 16.9\\
        &Qwen2.5-3B-SFT & 17.6 & 15.6 & \textbf{14.6} & \textbf{36.0}\\
        &Qwen2.5-7B-SFT & 23.5&16.8& 14.1 & 31.1\\

        \bottomrule
    \end{tabular}
    \caption{The complete scaling analysis of generation and citation labeling performance. The best performance is \textbf{bold}.}
    \label{tab: comparison_of_different_size}
\end{table*}

\begin{table*}[h]
\centering
\begin{tabular}{c p{12cm}}
\toprule
\multirow{7}{*}{\textbf{Instruction}} & Given the keyword chain of the question and its response, generate the original question. If the response is not informative enough to help you reconstruct the question, please rely on the provided keyword chain to generate the question. The keyword chain consists of terms where each term is a more specific or detailed subset of the previous one, with the last term being the most specific or important. The question you generate should focus on the last keyword in the chain and include it explicitly. \\  \midrule

\multirow{9}{*}{\textbf{Input}} & Given the following keyword chain and response: \\
& \textbf{Keyword Chain:} 72nd Primetime Emmy Awards, ceremony information, category and rule changes  \\
& \textbf{Response:} Several rule changes were announced in December 2019. first, episodes that were scheduled to air after the eligibility period closed\dots \\
&(Now, you should give me the original question given the keyword chain and its response. The output format should always be: Question: \$Question. Note that you should always try to generate it. Never ask for clarification or say you don't understand it in the generated question. Go ahead!) \\
\midrule

\multirow{2}{*}{\textbf{Model Output}} & \textbf{Question:} What were the category and rule changes for the 72nd Primetime Emmy Awards ceremony? \\
\bottomrule

\end{tabular}
\caption{An illustration of the prompt for question generation. The prompt consists of two parts: Instruction and Input.}
\label{prompt_of_generating_question}
\end{table*}

\afterpage{ 
\clearpage

\onecolumn
\centering
\begin{longtable}{c p{12cm}}

\endfirsthead
\endlastfoot

\toprule
\endhead
\bottomrule
\endfoot

\toprule
\multirow{12}{*}{\textbf{Instruction}} &  Given a topic and corresponding question and response pairs. 
\textcolor{red}{The questions are arranged in a logical, progressively deeper sequence, where each subsequent question delves deeper into the topic based on the previous one.} 
\textcolor{blue}{The questions are organized in a logical sequence where they are interconnected and may delve deeper into earlier topics, rather than following a direct, linear progression.} 
\textcolor{darkgreen}{The questions initially follow a logical progression but may shift to another topic as needed, reflecting a dynamic conversational flow rather than a strict linear order.}
I would like you to convert the original question into a conversational form. The goal is to rewrite it without any grammatical errors while preserving its original intent as closely as possible. It is necessary to consider the omission and reference to the previous question and response in the generated conversational question.\\  \midrule

\multirow{37}{*}{\textbf{Demonstration}} & I will give you one example multi-turn dialog, where each turn contains an original question, a conversational question, a response, and the corresponding analysis. \\ 
& \textbf{Example:} \\
& \textbf{Topic:} depression  \\
& \textbf{Conversations:} \\
& \textbf{Turn \#1:} \\
& \textbf{Original Question1:} What are the mechanisms of depression? \\
& \textbf{Conversational Question1:} What are the mechanisms of depression? \\

& \textbf{Response1:} The major neurotransmitters are acetylcholine, norepinephrine, dopamine, and serotonin. Many experts believe that an imbalance among the different neurotransmitters is the cause of depression. \\
& \textbf{Analysis:} The initial question addresses the mechanisms of depression, which leads to an explanation of neurotransmitters and their role in depression. \\

& \textbf{Turn \#2:} \\
& \textbf{Original Question2:} What is the role of serotonin in depression? \\
& \textbf{Conversational Question2:} What is the role of serotonin? \\

& \textbf{Response2:} When it comes to biological factors for depression, it is commonly accepted that serotonin plays a crucial role. In the brain, this neurotransmitter acts as a relay, transporting messages from one area to another. Given this critical task, an imbalance in serotonin levels may influence our emotions.  \\
& \textbf{Analysis:} Since "depression" was already mentioned in the previous turn, it's redundant to repeat it in the subsequent question. Therefore, by omitting "in depression", the question flows more smoothly and maintains coherence in the conversation. \\

& \textbf{Turn \#3:} \\
& \textbf{Original Question3:} How can I balance serotonin? \\
& \textbf{Conversational Question3:} So how can I balance it? \\

& \textbf{Response3:} You can create your own dopamine rewards by setting a goal, and then take small steps toward it every day. This can be starting a new exercise program, learning French, or challenging yourself to drive home from work a different way every day, preferably without the use of your GPS. \\
& \textbf{Analysis:} The use of "it" in the question effectively refers back to "serotonin" from the question in the previous turn. This helps maintain coherence in the conversation and avoids repetitive expressions. And we use "so" as a connecting word to make the conversation look smoother.\\

& \textbf{Turn \#4:} \\
& \textbf{Original Question4:} What foods boost dopamine? \\
& \textbf{Conversational Question4:} What foods boost it? \\

& \textbf{Response4:} Bananas, especially ripe bananas, are an exceptional food for regulating dopamine because they have a high concentration of tyrosine. Other foods that increase dopamine through the conversion of phenylalanine to tyrosine include almonds, apples, watermelons, cherries, yogurt, beans, eggs, and meats.  \\
& \textbf{Analysis:} Continuing the discussion on depression, the question asks about foods that boost dopamine, with "it" referring back to "dopamine" from the previous response.  \\
& \dots \\

\midrule

\multirow{38}{*}{\textbf{Input}} & \textcolor{orange}{Now I will give you a topic and the corresponding question and response pairs:}
\textcolor{darkgreen}{Now I will give you the question and response pairs:}\\
& \textcolor{orange}{\textbf{Topic:}} Solve Education  \\
\\
&\textbf{Id:} 1 \\
&\textbf{Question:} What is Solve Education and what are its main objectives? \\
&\textbf{Response:} solve education is a global technology non-profit organization founded in 2015 by peng tsin ong and Janine teo\dots\\
&\textbf{Id:} 2\\
&\textbf{Question:} What is the history of the organization Solve Education! since its inception? \\
&\textbf{Response:} solve education was founded in December 2015 after peng tsin ong attended the Africa Innovation Summit in Cape Verde in 2014   where he led a plenary session on financing innovation\dots\\
&\textbf{Id:} 3\\
&\textbf{Question:} What is the purpose of the Dawn of Civilization game developed by Solve Education, and how does it utilize its services to benefit education in low-income countries? \\
&\textbf{Response:} Dawn of Civilization is a city-building game which is based on various subjects and itself contains mini-games.\dots\\
&\textbf{Id:} 4\\
&\textbf{Question:} What is Ed the Learning Bot and how does it contribute to the services provided by Solve Education? \\
&\textbf{Response:} It is an AI-based chatbot used by solve education! on telegram. it helps underserved young people to learn about any subject with the help of gamified pictures and text\dots\\
&\textbf{Id:} 5\\
&\textbf{Question:} What is the purpose of the Content+ service on the Solve Education platform, and how do educational content providers interact with it? \\
&\textbf{Response:} it is an open platform which facilitates educational content providers to add their content in order to meet the needs of beneficiaries\dots\\
&\textbf{Id:} 6\\
&\textbf{Question:} What is Learnalytics and how does it function within Solve Education services? \\
&\textbf{Response:} It is a tool developed by solving education! which helps measure the performance.  additionally, learning portal analytics, provides context, the learning progress, and the performance of the students as they learn\dots\\
&\dots\\
\\
&(Please consider the question and response from the previous text when generating the current conversational question, but there is no need to generate the response. The output should be in the following format: \\
&Turn \#\$turn\_id \\ 
&Original Question: {\$original question} \\
&Conversational Question: {\$conversational question} \\ 
&Reason: {\$reason}) \\
\midrule

\multirow{34}{*}{\textbf{Model Output}} & \textbf{Turn \#1}\\
&\textbf{Original Question:} What is Solve Education! and what are its main objectives?\\
&\textbf{Conversational Question:} What is Solve Education and what are its main objectives?\\
&\textbf{Reason:} This is the first question in the series, so it introduces the topic and organization directly, laying the foundation for the conversation. No reference to prior questions is needed.\\
& \textbf{Turn \#2}\\
&\textbf{Original Question:} What is the history of the organization Solve Education since its inception?\\
&\textbf{Conversational Question:} Could you share more about the history of Solve Education since it started?\\
&\textbf{Reason:} The conversational question transitions smoothly from the introduction of the organization by asking for historical details. The phrase "since it started" connects directly back to the organization introduced in Turn \#1.\\
& \textbf{Turn \#3}\\
&\textbf{Original Question:} What is the purpose of the Dawn of Civilization game developed by Solve Education, and how does it utilize its services to benefit education in low-income countries?\\
&\textbf{Conversational Question:} Can you tell me about the Dawn of Civilization game and how it benefits education in low-income areas?\\
&\textbf{Reason:} This question builds on the previous introduction of Solve Education's initiatives. The phrase "the Dawn of Civilization game" is specific and follows logically from learning about the organization's history and broader objectives.\\
& \textbf{Turn \#4}\\
&\textbf{Original Question:} What is Ed the Learning Bot and how does it contribute to the services provided by Solve Education?\\
&\textbf{Conversational Question:} What role does Ed the Learning Bot play in Solve Education's services?\\
&\textbf{Reason:} The question uses "Ed the Learning Bot" directly and asks about its role, making it clear and focused on this specific service. It follows naturally after discussing another tool (Dawn of Civilization game) used by the organization.\\
&\dots\\
\bottomrule

\caption{An illustration of the prompt for generating conversational questions consists of three parts: Instruction, Demonstration, and Input. 
The \textcolor{red}{red section} is designated for LDS prompting, the \textcolor{blue}{blue section} for SIDS and STRW prompting, the \textcolor{darkgreen}{green section} for DTRW prompting, and the \textcolor{orange}{orange section} for LDS, SIDS, and STRW prompting.}
\label{prompt_of_generating_conversational_question}
\end{longtable}
\clearpage 
}

\clearpage
\begin{table*}[http]
\centering
\begin{tabular}{c p{12cm}}
\toprule
\multirow{3}{*}{\textbf{Instruction}} & Given the current question and the previous conversation history, summarize the conversation history. The summary should contain relevant information to help the agent provide a more informative response to the current question. \\  \midrule

\multirow{28}{*}{\textbf{Demonstration}} & I will give you several example dialogs, where each example contains the current question, the conversation history, the summarization, and the reason of generating such summarization. \\ 
& \textbf{Example \#1:} \\
& \textbf{Current Question:} What about their impact on battery longevity?  \\
& \textbf{Question1:} What are some of the key advancements in electric vehicle technology lately? \\
& \textbf{Response1:} There are improvements in battery technology for longer ranges, faster charging methods, and the integration of solar panels to help extend range. \\
& \textbf{Question2:} Do any particular models feature these solar panels? \\
& \textbf{Response2:} Yes, a number of new sedans and SUVs have solar roofs which can significantly increase daily driving range.\\
& \textbf{Question3:} What about the charging stations? Are they getting better too? \\
& \textbf{Response3:} Indeed, ultra-fast charging stations are now available that can boost a battery to 80\% in as little as 20 minutes.\\
& \textbf{Summarization:} The dialogue has highlighted advancements in electric vehicle technology, especially in battery efficiency, solar panel integration, and faster charging solutions. The conversation has particularly focused on the introduction of ultra-fast charging stations and briefly touched upon their efficiency in reducing charging times.\\
& \textbf{Reason:} The summary efficiently condenses prior discussions to provide context specifically for the user's current question about the impact of ultra-fast charging on battery longevity.\\
& \dots \\

\midrule

\multirow{12}{*}{\textbf{Input}} & Given the following question and its conversation history: \\
& \textbf{Current Question:} What are some other applications of this method, particularly in solving constrained optimization problems?  \\
& \textbf{Conversation History:} \\

& \textbf{Question1:} Can you explain what the multiplicative weight update method is and how it's used in decision making and prediction? \\
& \textbf{Response1:} The multiplicative weights update method is an algorithmic technique most commonly used for decision making and prediction\dots \\
&(Now, you should give me the summarization given the current question and its conversation history. The output format should always be: ``Summarization: \$Summarization''. Note that you should always try to generate it. Never ask for clarification or say you can't generate the summarization. Go ahead!) \\
\midrule

\multirow{6}{*}{\textbf{Model Output}} & \textbf{Summarization:} The conversation has detailed the multiplicative weight update method, describing its primary use in decision making and prediction, particularly through the example of prediction from expert advice. It also touched on its wide applications in fields like machine learning, optimization, theoretical computer science, and game theory, which relates to its potential in solving constrained optimization problems as queried in the current question. \\
\bottomrule

\end{tabular}
\caption{An illustration of the prompt for generating a conversation summary. The prompt consists of three components: Instruction, Demonstration, and Input.}
\label{prompt_of_summarizing_conversation_history}
\end{table*}

\begin{table*}[hb]
\centering
\begin{tabular}{c p{12cm}}
\toprule
\multirow{10}{*}{\textbf{Instruction}} & Based on the provided passages, 
\textcolor{red}{please answer the current question while considering the context of the conversations history. If the current question lacks complete information or contains omissions, use the conversation history to enhance understanding and provide a response.} 
\textcolor{blue}{please answer the question.}
\textcolor{darkgreen}{please answer the current question considering the summarization of the conversation history please. If the current question lacks complete information or contains omissions, use the summarization of conversation history to enhance understanding.}
Support your answer with citations in brackets. If the provided passages are not informative, please generate the answer based on your own knowledge. Don't say you can't answer the question.\\  

\midrule

\multirow{5}{*}{\textbf{Input}} 
& \textcolor{orange}{\textbf{Current Question:} \dots}\\
& \textcolor{blue}{\textbf{Question:} \dots}\\
& \textcolor{red}{\textbf{Conversation History:} \dots} \\
& \textcolor{darkgreen}{\textbf{Summarization of Conversation History:} \dots} \\
&\textbf{Passages:} \dots \\

\bottomrule

\end{tabular}
\caption{The prompt template of generating the response. The prompt consists of two parts, i.e., Instruction, and Input.
\textcolor{red}{red part} is for Raw Context and Last Response strategy prompting. 
The \textcolor{blue}{blue part} is for Rewrite Strategy prompting. 
The \textcolor{darkgreen}{green part} is for LLM Summarization Strategy prompting.
The \textcolor{orange}{orange part} is for Raw Context, Last Response Strategy, and LLM Summarization Strategy prompting.}
\label{prompt_of_generating_reseponses_with_citation_labeling}
\end{table*}

\clearpage
\begin{table*}[http]
\centering
\begin{tabular}{p{14cm}}
\toprule
\textbf{Question1:} What were the key details and outcomes of the 72nd Primetime Emmy Awards?\\
\textbf{Response1:} The 72nd Primetime Emmy Awards honored the best in American prime time television programming from June 1, 2019, until May 31, 2020, as chosen by the Academy of Television Arts \& Sciences.[65215]\dots\\
\textbf{Golden Retrieval Passages IDs:} 65215\\
\textbf{Golden Rewrite:} What were the key details and outcomes of the 72nd Primetime Emmy Awards?\\
\textbf{URL}: \url{https://en.wikipedia.org/wiki/72nd_Primetime_Emmy_Awards}\\

\\

\textbf{Question2:} Could you tell me about the winners and nominees from this event?\\
\textbf{Response2:} The nominations for the 72nd Primetime Emmy Awards were announced on July 28, 2020, by host Leslie Jones and presenters Laverne Cox, Josh Gad, and Tatiana Maslany.[65226]\dots\\
\textbf{Golden Retrieval Passage IDs:} 65216, 65217, 65218, 65219, 65220, 65221, 65222, 65223, 65224, 65225, 65226, 65227, 65228, 65229, 65230, 65231, 65232\\
\textbf{Golden Rewrite:} Who were the winners and nominees at the 72nd Primetime Emmy Awards?\\
\textbf{URL}: \url{https://en.wikipedia.org/wiki/72nd_Primetime_Emmy_Awards}\\

\\

\textbf{Question3:} Who received the Governors Award and why were they chosen?\\
\textbf{Response3:} The Governors Award was presented to Tyler Perry and the Perry Foundation ``in recognition of their unparalleled contributions to shaping the television medium'' and for their ``inclusion, engagement, employment and other philanthropic initiatives''.[65233][65234]   The award was moved to the main telecast from its usual presentation at the Creative Arts Emmys.[65235]\\
\textbf{Golden Retrieval Passage IDs:} 65233, 65234, 65235\\
\textbf{Golden Rewrite:} Who received the Governors Award at the 72nd Primetime Emmy Awards and for what reasons?\\
\textbf{URL}: \url{https://en.wikipedia.org/wiki/72nd_Primetime_Emmy_Awards}\\

\dots \\
\dots \\
\dots \\
\textbf{Question10:} Who performed during the in Memoriam Segment, and which notable figures were mentioned?\\
\textbf{Response10:} The annual in Memoriam Segment featured H.E.R. performing ``Nothing Compares 2 U'' on piano and electric guitar .[65284][65285]\dots\\
\textbf{Golden Retrieval Passage IDs:} 65284, 65285, 65286, 65287\\
\textbf{Golden Rewrite:} Who performed during the in Memoriam Segment at the 72nd Primetime Emmy Awards, and which notable figures were mentioned?\\
\textbf{URL}: \url{https://en.wikipedia.org/wiki/72nd_Primetime_Emmy_Awards}\\

\bottomrule

\end{tabular}
\caption{A conversation from CORAL.}
\label{example_of_coral}
\end{table*}

\end{document}